# Estimating the quality of published medical research with ChatGPT[1]

Mike Thelwall
Information School, University of Sheffield, UK. https://orcid.org/0000-0001-6065-205X
Xiaorui Jiang
Information School, University of Sheffield, UK. https://orcid.org/0000-0003-4255-5445
Peter A. Bath
Health Informatics Research Group, Information School, University of Sheffield, UK. https://orcid.org/0000-0002-6310-7396

Estimating the quality of published research is important for evaluations of departments, researchers, and job candidates. Citation-based indicators sometimes support these tasks, but do not work for new articles and have low or moderate accuracy. Previous research has shown that ChatGPT can estimate the quality of research articles, with its scores correlating positively with an expert scores proxy in all fields, and often more strongly than citation-based indicators, except for clinical medicine. ChatGPT scores may therefore replace citation-based indicators for some applications. This article investigates the clinical medicine anomaly with the largest dataset yet and a more detailed analysis. The results showed that ChatGPT 4o-mini scores for articles submitted to the UK's Research Excellence Framework (REF) 2021 Unit of Assessment (UoA) 1 Clinical Medicine correlated positively (r=0.134, n=9872) with departmental mean REF scores, against a theoretical maximum correlation of r=0.226. ChatGPT 4o and 3.5 turbo also gave positive correlations. At the departmental level, mean ChatGPT scores correlated more strongly with departmental mean REF scores (r=0.395, n=31). For the 100 journals with the most articles in UoA 1, their mean ChatGPT score correlated strongly with their REF score (r=0.495) but negatively with their citation rate (r=-0.148). Journal and departmental anomalies in these results point to ChatGPT being ineffective at assessing the quality of research in prestigious medical journals or research directly affecting human health, or both. Nevertheless, the results give evidence of ChatGPT's ability to assess research quality overall for Clinical Medicine, where it might replace citation-based indicators for new research.

# Introduction

Research quality evaluation is important for departmental evaluations and academic career decisions. Unfortunately, the evaluators may not have time to fully read the work assessed and may instead rely on the reputation or Journal Impact Factor of the publishing journals, on the citation counts for individual articles, or on the reputation or career citations of the author. Whilst journal-based evidence is not optimal (Waltman & Traag, 2021), the main article-level indicator, citation counts, only directly reflects the scholarly impact of work and not its rigour, originality, and societal impacts (Aksnes, et al., 2019), all of which are relevant quality dimensions (Langfeldt et al., 2020). Moreover, article citation counts are ineffective for newer articles (Wang, 2013). In response, attempts to use Large Language Models (LLMs) to estimate the quality of academic work have shown that ChatGPT quality scores are at least as effective as citation counts in most fields and substantially better in a few (Thelwall & Yaghi, 2024). Clinical medicine is an exception, however, with ChatGPT research quality scores having a small negative correlation with the mean scores of the submitting department in the Research Excellence Framework (REF) Clinical Medicine Unit of Assessment (UoA) (Thelwall, 2024, 2025; Thelwall & Yaghi, 2024). It is therefore important to find the reason for this anomaly and, if possible, create an effective LLM-based research quality assessment method for clinical medicine.

---
[1] Thelwall, M., Jiang, X., & Bath, P. (2025, to appear). Estimating the quality of published medical research with ChatGPT. Information Processing & Management.

The goal here is not to evaluate articles in the sense that a reviewer or a human expert might, but only to provide a research quality estimate that can be used when an evaluation is not required or impossible (e.g., because there are too many outputs to be assessed by a human or there are no available experts). These situations might include busy appointment panels assessing the promise of many applicants partly based on the publications on their CVs (Ghani, 2020), situations where two experts disagree on the value of an article and consult citation rates as a tiebreaker (Wilsdon et al., 2015), and larger scale evaluations of departments, universities or countries where the cost of formal evaluations of outputs would be higher than the value of the information generated, such as for the Publications section of the State of US Science and Engineering reports (e.g., ncses.nsf.gov/pubs/nsb20243).

Although ChatGPT is now widely used to support academic research (Eppler et al., 2024; Owens, 2023), and it has been systematically evaluated for natural language processing tasks like question answering, sentiment analysis, and text summarisation (Kocoń et al., 2023) it has sometimes been applied to academic text processing tasks where a score must be given to a document based on complex criteria. Examples of this include pre-publication peer review recommendations (Zhou et al., 2024; Saad et al., 2024) and post-publication expert review quality scoring (Thelwall & Yaghi, 2024) as well as scoring impact case studies for reach and impact (Kousha & Thelwall, 2024). These show that ChatGPT works well with system instructions like those given to human experts, presumably because it is partly trained this way (Ouyang et al., 2022). It also seems to have a degree of effectiveness at translating authorial claims in abstracts into reasonable scores, perhaps also considering some wider context. For optimal results, the default parameters for the ChatGPT API work well and varying them may not help (Thelwall, 2025). Nevertheless, scores can be improved by submitting the same query multiple times and averaging the results (Thelwall, 2024). This seems to be a way of leveraging ChatGPT's internal probability distribution, which reveals the level of confidence of ChatGPT's model in the scores within its reports.

For wider context, there seems to have been only one previous attempt to systematically provide post publication research quality scores for medical research, in the form of the Faculty Opinions system (formerly, Faculty of 1000, F1000, and F1000Prime) that gives expert post-publication scores (Chen et al., 2024) and tags (e.g., "Controversial", "Good for Teaching") to biomedical articles (Wang et al., 2020). An investigation into four medical journals found that higher rated articles tended to be more cited in three of the journals (Wang et al., 2020). The scores are available only to subscribers and probably only for a small fraction of biomedical research, however.

Given the promise shown by ChatGPT for a variety of evaluation-type tasks, as discussed above, and the positive results previously found for all fields except UoA 1 Clinical Medicine (Thelwall & Yaghi, 2024), the motivation of this study is to investigate whether ChatGPT can give useful score estimates for medical research by checking the previous Clinical Medicine findings using a larger dataset and investigating the reasons for the negative correlation. A methodological limitation of the previous study is that it sampled articles only from high and low scoring departments and if any of these had an unusual publishing strategy then this could influence the findings. The current paper therefore seeks a more comprehensive evaluation of UoA 1 as well as follow-up investigations to identify reasons for weak or negative correlations. In addition to article-level comparisons (RQ1) (as previously reported: Thelwall & Yaghi, 2024), departmental-level comparisons (RQ2) are useful because this is the level at which REF quality data is available and used in practice. This also corresponds to the departmental evaluations that citation-based indicators are sometimes used for. Journal-level comparisons (RQ3) may give a different perspective and, since most well-known indicators for journals use citation rates rather than average quality, this comparison is included.

- RQ1: Do ChatGPT scores correlate positively with REF scores for articles in UoA 1 Clinical Medicine? As discussed in the methods, REF scores are not available for articles, so departmental mean REF scores are used as a proxy for article quality.
- RQ2: Do departmental mean ChatGPT scores correlate positively with departmental mean REF scores for departments in UoA 1?
- RQ3: Do journal mean ChatGPT scores correlate positively with (a) the mean of the departmental mean REF scores and (b) the average citation rate for the journal's articles in UoA 1?

RQ4: Which types of clinical medicine articles tend to get high and low scores from ChatGPT?

## Methods

The research design was to obtain ChatGPT scores for as many as possible of the articles submitted to REF2021 in UoA 1 Clinical Medicine and then compare these scores with departmental REF mean for journal articles individually, by department, and by journal. The individual REF scores for journal articles are not known because only the number of articles generating each quality score is reported publicly, and the departmental REF mean is the best available quality score proxy.

*Data*

The data used is from the UK Research Excellence Framework, which is a periodic national evaluation of the quality of academic research outputs, activities and impacts over the previous seven years. For REF2021, the scores given by 1120 experts grouped into 34 field-based Units of Assessments, determined the destination of the annual £2 billion block research grants to UK universities. As part of this, the experts individually scored 185,594 research outputs (mostly journal articles) for research quality (rigour, originality and significance) on a four-point scale: 1* recognised nationally; 2* recognised internationally; 3* internationally excellent; 4* world leading. Each output was independently scored by at least two experts, then all scores were ratified by UoA subpanels after additional norm referencing between UoAs. These scores were not released and have been destroyed but list of outputs and departmental score profiles (see below) are publicly available.

The REF outputs are available in a spreadsheet that can be downloaded from the REF website (results2021.ref.ac.uk). This set was filtered to exclude everything except the journal articles in UoA 1. These were then matched with corresponding records in Scopus by DOI, when available. Scopus records were needed to access the abstracts and citation counts of these articles, which were not in the REF data. This gave a set of 9,905 journal articles from UoA 1 matched with Scopus records. After excluding articles without abstracts, 9,872 remained for the main analysis. ChatGPT needs abstracts to score articles with the methods used here.

**Departmental REF scores for articles**: Each university submission to UoA 1 is called a "department" here although it may not map to an organisational unit or units with this name. Unfortunately, the REF scores for individual articles were never made public and were systematically deleted before the aggregate results were published. Instead, public information about departmental scores can be used to calculate departmental mean REF scores and these departmental scores can be used as approximate estimates of the REF quality of each of a department's articles. This proxy is not ideal but seems to be the only method to get any kind of quality estimate score for the articles. Unless there is a substantial departmental bias within ChatGPT, a positive correlation between ChatGPT scores and REF scores at the article level should translate into a weaker correlation between ChatGPT scores and departmental average REF scores at the article level.

For the departmental average REF score calculation, the departmental results could be downloaded from the official website (results2021.ref.ac.uk). For each department, the percentage of outputs scoring each one of the four quality levels (1*, 2*, 3* or 4*) is recorded and these percentages could be used to calculate the departmental mean REF scores. This is imperfect, however, because some journal articles are not in Scopus, some departmental outputs are not journal articles and some journal articles were deemed out of scope and not given a REF score. Thus, instead we used the departmental average from the valid journal article in Scopus submitted by each department. This information is not public but had been calculated for a previous project before the individual article level scores were deleted. Thus, the departmental REF averages used here were calculated from all valid Scopus-indexed articles from UoA 1 (unfortunately including the few without abstracts), and this is used as the REF score estimate for the articles.

Following previous studies (Thelwall, 2024, 2025, Thelwall & Yaghi, 2024) only the titles and abstracts were used for input and not the full text. There are multiple reasons for this: previous research has given better results from titles and abstracts than from full texts from ChatGPT (Thelwall, 2024, 2025); in the UK it is a breach of copyright to use full texts for machine learning without permission, except for academic research, so full-text analyses do not have practical applications; the research team did not have access to the full text of all 9,872 articles analysed, nor a facility to download them automatically; ChatGPT does not allow automatic uploading of documents (at the time of the experiments), so only the text extracted from the PDFs could be uploaded, without images and formulae, which would make them confusing documents; and uploading full texts makes each query more expensive, reducing the practicality of the method (the experiments in the current paper cost about $500 in ChatGPT usage charges).

**ChatGPT Scores**: Each of the 9,872 articles were submitted to ChatGPT 4o-mini (version gpt-4o-mini-2024-07-18) to obtain a score. Only the title and the abstract were submitted (with the prompt, "Score this: title\nAbstract\nabstract") and the REF Panel A guidelines (i.e., the guidelines relevant to clinical medicine) were used as the system instructions for ChatGPT (as in a previous article: Thelwall & Yaghi, 2024). ChatGPT's output is a report almost always containing a recommended score, and these scores were extracted by a series of text processing rules in Webometric Analyst (AI menu, Extract scores from ChatGPT reports option: github.com/MikeThelwall/Webometric_Analyst), with human input when the rules failed. Each article was submitted five times to ChatGPT, and the mean value used as the article's ChatGPT score. As previously shown, ChatGPT's scores can vary for the same prompt and this averaging process improves the accuracy of the score. About five times is enough before the additional iterations add a relatively small amount to the accuracy (Thelwall & Yaghi, 2024). Inputting legally accessed texts into artificial intelligence systems for academic research is legal in the UK and does not require copyright holder permission (Bristows, 2023). The ChatGPT API does not learn from the data ingested (OpenAI, 2024) and so there is no possibility of secondary copyright infringement.

For secondary tests of model accuracy, the above was repeated for the full model ChatGPT 4o (gpt-4o-2024-11-20) and the earlier and smaller model ChatGPT 3.5 (gpt-3.5-turbo-0125).

No fine tuning was used to contrast with the above zero-shot approach. This is because no article REF scores are known and so there is no training data. The evaluation approach (described below) uses an indirect approach to estimate accuracy without knowing the correct scores. Whilst it would be possible to leverage this indirect data to generate score estimates for articles to try a few-shot approach, the indirect data is of a different type (not integers) and because it is tied to departments and they can have specialties, this could improve accuracy by "cheating" through learning the topics of the stronger departments.

Hence it would not be able to fairly assess the level of improvement achievable by any fine tuning.

Although the ChatGPT API allows parameters to be set that influence the output, the default settings were used (temperature = 1; top p = 1) because these worked best a previous investigation of the influence of the parameters for the same task. This previous study assessed temperature in {0.1,0.5, 1, 1.5, 2}, and top p in {0.25, 0.5, 0.75, 1} (Thelwall, 2025). For example, lower temperatures decreased the variety that the averaging process takes advantage of (e.g., always scoring 3* rather than usually scoring 3* and occasionally scoring 4*, revealing that ChatGPT "considered" the article to be a high 3*), whereas higher temperatures led to reports that diverged from the task.

**Departmental and journal ChatGPT score**: This is the mean of all ChatGPT scores for all articles associated with the department/journal from the 9,872 analysed.

Although the Scopus data includes citation counts, as needed for RQ3, they need processing to be more useful because the articles are from different years (2014 to 2020) and fields (mostly related to Medicine, but not all) and citation rates vary naturally between fields and years. Sets of citation counts are also highly skewed, so a log transformation is needed before any averaging is applied. Thus, each citation count was transformed with Log(1+x) to reduce skewing, then divided by the mean of the Log(1+x) values for all articles published in Scopus in the same Scopus narrow field and year (including articles not submitted to the REF). The result is the Normalised Log-transformed Citation Score (NLCS), which is fair to compare between articles published in different fields and years (Thelwall, 2017).

**Citation rates for journals**: The citation rate for each journal was calculated as the mean of the NLCS for all the articles associated with that journal from the 9,872 analysed. This is known as the Mean NLCS (MNLCS). This definition excludes all non-REF articles by design (these are only used for the normalisation procedure). Thus, even though the journal MNLCS is a bit like the Journal Impact Factor (JIF), it is based on a different set of articles.

## Analysis

The primary statistical test was the Pearson correlation coefficient. This is more informative than a direct measure of accuracy (e.g., mean absolute deviation) because ChatGPT's score estimates can easily be scaled with a transformation. Pearson correlations were chosen in preference to Spearman because the data are not highly skewed, and Pearson is finer grained. Although some of the data analysed derives from ranks, the numbers correlated are not ranks and are often not integers due to averaging.

For RQ4, a Word Association Thematic Analysis (Thelwall, 2021) was used to identify themes related to articles with high ChatGPT scores by analysing the words found disproportionately often in the articles scoring at least the median ChatGPT score or more (ChatGPT average >= 3.6 n=5353) compared to articles with a below-median ChatGPT score (ChatGPT average <3.6, n= 4637). This method is conservative and statistically-oriented, designed to give evidence-based themes in sets of texts. For this, article titles and abstracts were fed into the text analytics software Mozdeh (github.com/MikeThelwall/Mozdeh) together with their ChatGPT scores and it was used to construct a list of terms occurring (statistically significantly) disproportionately often in the higher ChatGPT score set (for details see: Thelwall, 2021). The top 25 words were examined to find their typical context in the articles. Finally, the term contexts were clustered reflexively into themes by the first author. This was repeated for the lower scoring articles.

# Results

## *RQ1: Article-level analysis*

The Pearson correlation between an article's ChatGPT 4o-mini score (i.e., the mean of 5 scores per article) and the REF score of its submitting department is 0.134 (95% confidence interval: 0.114, 0.153, n=9872). This is weak but positive and statistically significantly greater than 0, giving a positive answer to the first research question. The theoretical maximum correlation for this variable is the Pearson correlation between the individual article REF scores and the departmental mean values which can be calculated from the score distributions without knowing the individual article scores, and is 0.226 (at the $R^2$ level, the 0.134 correlation is 35% of the maximum). Thus, the underlying correlation between the ChatGPT article is likely to be substantially higher and may be moderate rather than weak.

The ChatGPT scores tend to be above the departmental REF mean values, with the ChatGPT overall mean being 3.55 and the mean REF score (per article rather than per department) being 3.27. Thus, ChatGPT might be effectively rounding up scores between 3* and 4* to 4* that REF assessors rounded down to 3*.

The Pearson correlation between ChatGTP 3.5 turbo scores (mean of 5 per article) is lower at 0.123 (95% confidence interval: 0.103, 0.142, n=9872) with a lower mean score (3.17) than 4o-mini. The Pearson correlation between ChatGTP 4o scores (mean of 5 per article) is higher at 0.153 (95% confidence interval: 0.134, 0.173, n=9872), with a slightly higher mean score (3.56). At the $R^2$ level, the 0.153 correlation is 46% of the maximum, following the same logic as above. For all three models, the correlation is substantially higher for the average of the five ChatGPT iterations than for the individual sets of ChatGPT scores (47% higher for ChatGPT 3.5 turbo, 22% higher for 4o-mini, and 29% higher for 4o). The lower correlation for the older model and the higher correlation for the full 4o model are unsurprising. The full 4o model can be expected to retain more information from the training corpus texts that it has read than the smaller and cheaper 4o-mini and therefore to make more precise associations from new prompts. The remainder of this section focuses on the 4o-mini results.

In an attempt to bring the ChatGPT 4o-mini average closer to the REF average score, a number of different system prompting strategies were developed to encourage ChatGPT to be stricter, by explicitly telling it to be "strict", "very strict", or "draconian". These were tested on a tiny development set of 20 articles from The Lancet and eLife from 2024. The only strategy found that seemed to be effective at reducing the average scores was to change the start of the system instructions from "You are an academic expert, assessing academic journal articles" to "You are a draconian academic expert, harshly assessing academic journal articles", and allowing it to use half points by adding, "Use half points if a study is between two scores." Because of the issues mentioned in the RQ3 results below, the following was also added, "Research that directly affects human health is particularly valuable." The experiment (submitting all 9,872 articles to ChatGPT 4o-mini five times and taking the average) was repeated with this revised prompt and the average ChatGPT score reduced slightly from 3.55 to 3.47 but this seemed to make the ChatGPT scores less informative because the correlation with departmental REF mean values reduced from 0.134 to 0.117. The original 4o-mini dataset alone was therefore used for the remainder of this paper.

## RQ2: Department-level analysis

If the mean ChatGPT score of all of a department's articles is correlated against its mean REF score for journal articles then the correlation is 0.395 (95% CI: 0.047, 0.657, n=31), which is higher than at the article level, as expected. Nevertheless, this correlation is only modest, and a scatter plot reveals the existence of two substantial outliers (Figure 1).

- Warwick University (78 articles) has the second lowest REF score and the second highest ChatGPT score.
- Leicester University (81 articles) has the highest REF score and the 6$^{th}$ lowest mean ChatGPT score.

The anomaly can be investigated by examining the journals mainly submitted by these two institutions. Whilst the most popular journals for Leicester where arguably the two most prestigious medical journals, the New England Journal of Medicine (NEJM, 13 articles) and The Lancet (10 articles), the single most popular journal for Warwick was eLife, a general biomedical and life sciences journal (11 articles). Thus, since the two departments seem to have different publishing patterns, the anomaly might be due to ChatGPT unduly favouring Warwick's topics or the journals that Warwick publishes in, relative to Leicester. The next section focuses on journals.

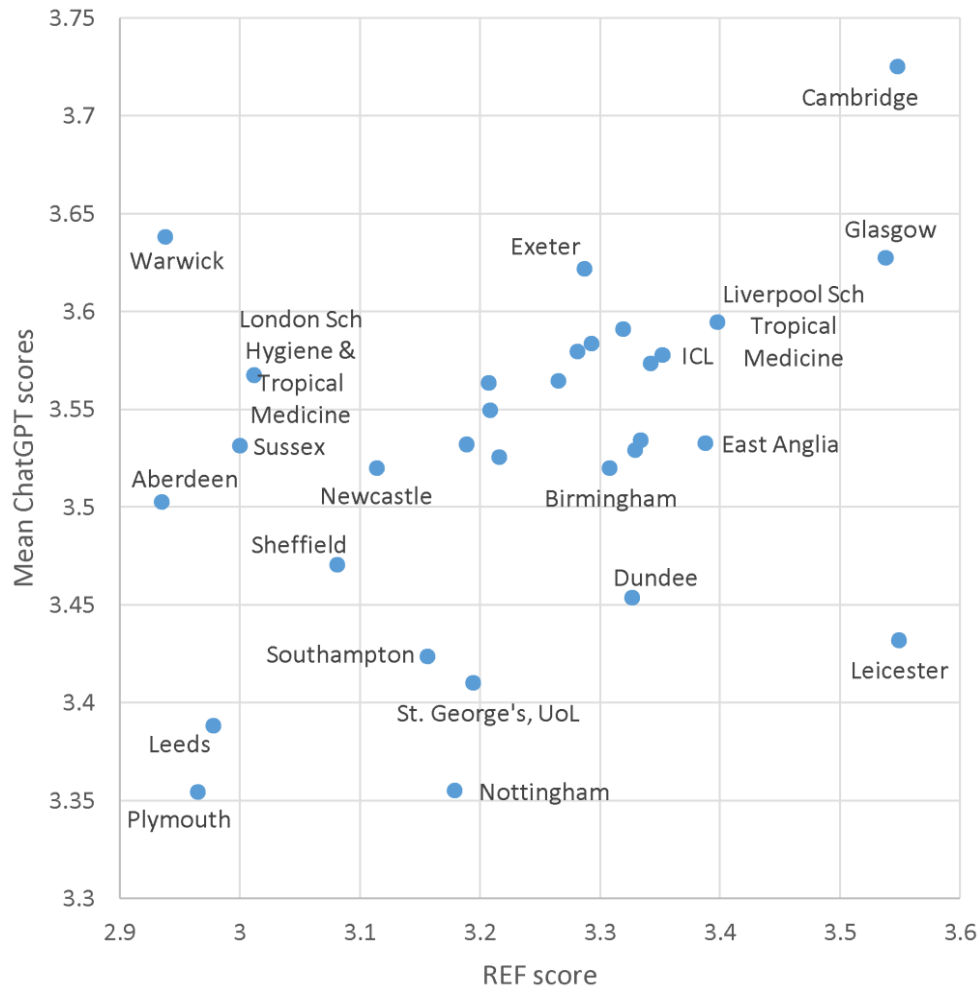

Figure 1. Mean ChatGPT scores for all of a department's REF articles against the departmental mean REF score.

*RQ3: Journal-level analysis*

Altogether, 767 different journals had articles in the UoA 1 sample. The top journals in terms of explaining the results are those with the most articles, however. Thus, correlations were calculated overall and just for the top journals, using different cut-offs since there is not a natural choice. Including smaller journals adds noise to the data and tends to reduce the correlation, whereas only considering the top journals would be unrepresentative.

For the 50 largest journals for UoA 1 there was a moderately strong (r=0.517, n=50) Pearson correlation between the journal REF score (the mean REF score of the departments of the articles in the journal) and the journal ChatGPT score (the mean ChatGPT score of articles in the journal) (Table 1). Although there is also a positive correlation between journal REF score and journal mean citation rates (r=0.245, n=50), there is a negative correlation (r=-0.245, n=50) between journal ChatGPT scores and journal mean citation rates (MNLCS). Thus, and unexpectedly, ChatGPT tends to give lower scores to articles in more cited journals in UoA 1.

Table 1. Pearson correlations [95% confidence intervals] between journal mean citation rate, journal mean departmental REF score, and journal mean ChatGPT score for the N journals with the most qualifying articles in UoA 1.

| Top N journals | REF vs GPT | GPT vs MNLCS | REF vs MNLCS |
|---:|---:|---:|---:|
| 50 | 0.517 [0.279, 0.695] | -0.245 [-0.490, 0.036] | 0.245 [-0.036, 0.490] |
| 100 | 0.495 [0.331, 0.630] | -0.148 [-0.335, 0.050] | 0.288 [0.097, 0.458] |
| 200 | 0.378 [0.253, 0.491] | -0.035 [-0.173, 0.104] | 0.305 [0.174, 0.426] |
| 500 | 0.232 [0.147, 0.313] | 0.022 [-0.066, 0.109] | 0.257 [0.173, 0.337] |
| 767 | 0.152 [0.082, 0.220] | 0.037 [-0.034, 0.108] | 0.149 [0.079, 0.217] |

A scatter plot of journal mean citation rates against journal mean ChatGPT scores shows that four well known prestigious medical journals are highly cited (as expected) but tend to receive low ChatGPT scores: NEJM, The Lancet, JAMA, and The BMJ. There are also many journals that get relatively high ChatGPT scores for their citation rates (bottom right corner of Figure 2). This issue is analysed in the Discussion. Figures 1 and 2 are also available online with comparable scales at https://doi.org/10.6084/m9.figshare.28378274.

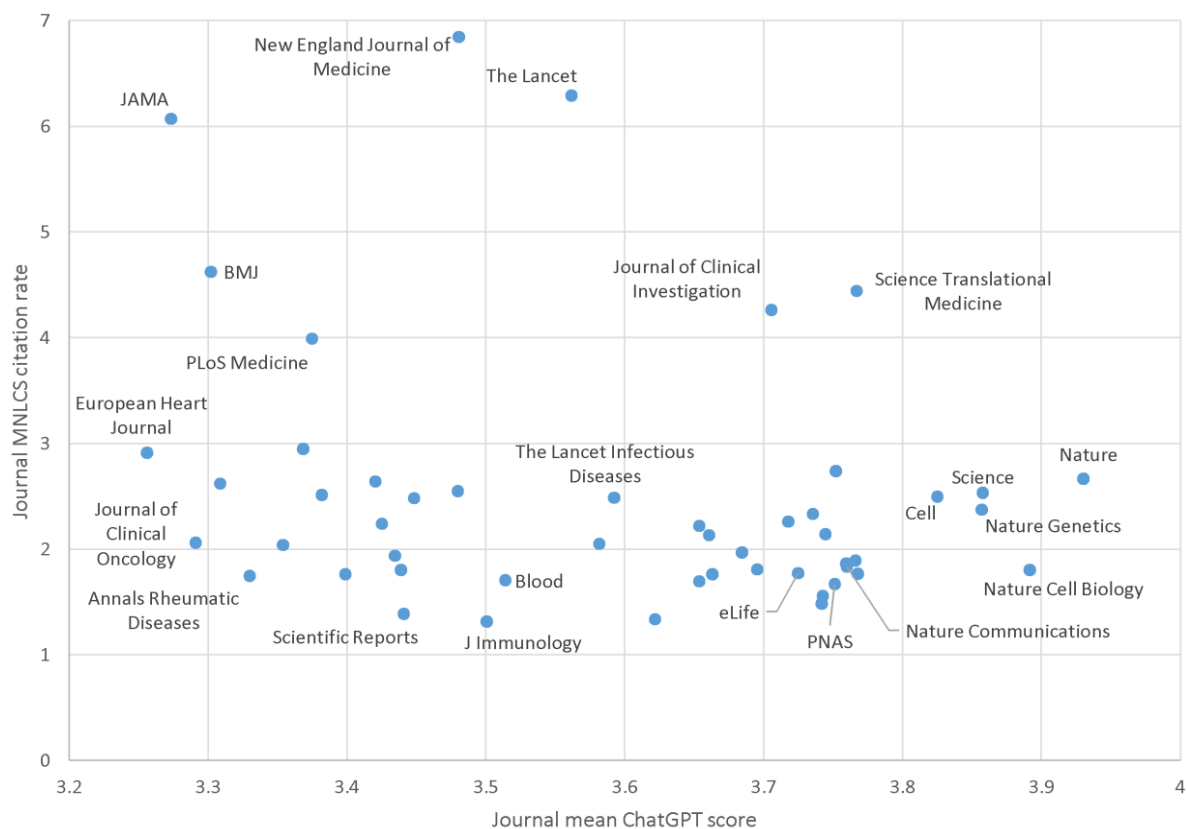

Figure 2. Mean logged citation rate (MNLCS) against mean ChatGPT scores for all of a journal's qualifying REF articles for the 50 journals with the most articles in UoA 1.

## RQ4: Types of articles with higher ChatGPT scores

The Word Association Thematic Analysis to find article types attracting above median or below median ChatGPT scores revealed several clear patterns. Themes found in the higher scoring articles (Table 2) include genetics (e.g., words like: genetic, genome-wide), style (e.g., here, we, show, that), exploratory/theoretical (e.g., mechanism, complex, drive, pathway). For example, the term "we" occurred in 92.7% of the abstracts given a higher score by ChatGPT but only 69.3% of articles given a lower score by ChatGPT. This was a writing style issue: due to phrases like "here we show that" tending to occur disproportionately often in higher scoring article abstracts. This style seems to be associated with prestigious medical journals so it is probably a side effect of higher quality research being published in journals that encourage this style rather than a direct effect of ChatGPT associating this style with higher quality research (although it may have learned that this style associates with higher quality research or prestigious journals).

In contrast, themes for the lower scoring articles (Table 3) include use of the past tense (e.g., were, was, had), structured abstract terms (e.g., methods, conclusion), patient/participants (e.g., patient, participant outcome, age), and statistics (e.g., p mean, ci).

Overall, this suggests that theoretical studies scored higher, perhaps by revealing more substantial results, whereas studies directly informing human health decisions scored lower. This is perhaps demonstrated most clearly by the terms "patient" and "participant" (suggesting a study directly related to human health) occurring more often in lower scoring articles and "human" (suggesting a wider perspective) occurring more often in higher scoring articles. In this context, the cell biology, genetics and molecular biology themes seem to be relatively theoretical.

Table 2. Words and associated themes in the titles and abstracts of articles attracting an average ChatGPT score of at least 3.6 (top 25). All differences are statistically significant with p<0.001 for a chi squared test after Benjamini–Hochberg familywise error correction.

| Word | Higher GPT | Lower GPT | Articles | Chisq | Theme |
|---|---|---|---|---|---|
| we | 92.7% | 69.3% | 8064 | 899.6 | Style/"Here we show that" |
| here | 45.2% | 17.6% | 3182 | 851.1 | Style/"Here we show that" |
| that | 81.4% | 61.9% | 7129 | 465.2 | Style/"Here we show that" |
| show | 32.6% | 14.5% | 2382 | 440.6 | Style/"Here we show that" |
| cell | 53.2% | 33.3% | 4325 | 395 | Cell biology |
| reveal | 15.7% | 3.9% | 1001 | 377.1 | Exploratory/theoretical |
| human | 35.0% | 19.5% | 2737 | 293.2 | Exploratory/theoretical (e.g., mice v.) |
| mechanism | 24.7% | 12.0% | 1847 | 260.4 | Exploratory/theoretical |
| gene | 30.2% | 16.7% | 2356 | 248.4 | Genetics |
| genome | 9.8% | 2.5% | 630 | 219.1 | Genetics |
| demonstrate | 19.1% | 9.1% | 1423 | 196.1 | Exploratory/theoretical |
| our | 31.6% | 19.2% | 2543 | 195.2 | Style/"Here we show that" |
| protein | 27.8% | 17.1% | 2249 | 158.5 | Molecular biology |
| genetic | 16.6% | 8.3% | 1253 | 151.3 | Genetics |
| genome-wide | 7.5% | 2.1% | 492 | 151 | Genetics |
| molecular | 13.4% | 6.1% | 984 | 148 | Molecular biology |
| complex | 13.3% | 6.0% | 977 | 146.8 | Exploratory/theoretical |
| mutation | 15.6% | 8.0% | 1184 | 134.2 | Genetics |
| genomic | 7.5% | 2.4% | 503 | 133 | Genetics |
| drive | 7.1% | 2.2% | 471 | 128.3 | Exploratory/theoretical |
| pathway | 18.40% | 10.60% | 1455 | 120.5 | Exploratory/theoretical |
| distinct | 10.50% | 4.60% | 761 | 120.1 | Genetics; Cell biology |
| through | 20.00% | 11.90% | 1596 | 119 | Cell biology |
| identify | 18.40% | 10.70% | 1461 | 116.1 | Genetics |
| sequencing | 9.80% | 4.20% | 708 | 114.7 | Genetics |

Table 3. Words and associated themes in the titles and abstracts of articles attracting an average ChatGPT score below 3.6 (top 25). All differences are statistically significant with p<0.001 for a chi squared test after Benjamini–Hochberg familywise error correction. A tilde ~ indicates a partial match for the stated theme.

| Word | Lower GPT | Higher GPT | Articles | Chisq | Theme |
|---|---|---|---|---|---|
| were | 64.9% | 34.5% | 4817 | 910.4 | Past tense |
| was | 67.0% | 36.6% | 5020 | 906.9 | Past tense |
| method | 47.0% | 22.2% | 3340 | 675.2 | Structured abstract |
| conclusion | 30.2% | 10.8% | 1967 | 578.9 | Structured abstract |
| objective | 22.9% | 6.6% | 1405 | 537.2 | Structured abstract |
| background | 33.6% | 15.8% | 2387 | 422.3 | Structured abstract |
| study | 44.7% | 25.0% | 3381 | 420.3 | Structured abstract |
| patient | 47.4% | 27.6% | 3643 | 414.8 | Patient/participant studies |
| outcome | 27.3% | 11.3% | 1856 | 408.8 | Patient/participant studies |
| result | 54.1% | 34.1% | 4295 | 397.4 | Patient/participant studies |
| trial | 24.3% | 9.5% | 1625 | 394.5 | Patient/participant studies |
| year | 25.4% | 10.6% | 1733 | 375.4 | Patient/participant studies |
| compared | 27.2% | 12.1% | 1893 | 359.9 | Patient/participant studies~ |
| no | 23.9% | 10.0% | 1630 | 346.8 | Patient/participant studies~ |
| ci | 19.8% | 7.7% | 1319 | 309.8 | Statistics |
| had | 26.0% | 12.3% | 1850 | 302.5 | Past tense |
| group | 22.7% | 10.0% | 1577 | 294.5 | Patient/participant studies~ |
| there | 19.2% | 7.5% | 1281 | 294.2 | Unknown |
| who | 19.3% | 7.9% | 1308 | 279.6 | Patient/participant studies~ |
| age | 17.6% | 6.8% | 1173 | 275.9 | Patient/participant studies |
| included | 14.30% | 4.60% | 907 | 274.9 | Patient/participant studies~ |
| mean | 13.60% | 4.40% | 861 | 263.4 | Statistics |
| p | 24.50% | 12.10% | 1770 | 253.9 | Statistics |
| month | 14.40% | 5.10% | 939 | 246.6 | Patient/participant studies |
| participant | 15.20% | 5.80% | 1013 | 235 | Patient/participant studies |

# Discussion

The results are limited by the nature of the sample used, which is from a single country and consists of the self-selected best work of the publishing academics. It may also exclude work from weaker departments if they chose to submit to a related UoA instead, or if individuals were submitted to a different UoA related to medicine (e.g., UoA2, Public Health, Health Services and Primary Care). The strength of the correlations reported is likely to differ between LLMs and between versions of an LLM. Finally, the positive correlations reported may well increase in strength for newer LLMs, although it is not clear if the negative correlations will become positive or stronger.

In terms of inputs, ChatGPT was used to assess the quality of articles based on their titles and abstracts, since previous research had found this input to give the best results (Thelwall, 2025), whereas REF panel members are expected to read and assess the whole paper. The current generation of LLMs can process full texts rather than just abstracts and titles but ChatGPT in the small-scale tests so far has given better results with just titles and abstracts. As argued above, copyright considerations seem to currently rule out the uploading of most full texts to public LLMs for fuller testing, although the legal situation for LLMs is currently unclear and the law may change soon. Titles and abstracts form a clear legal exception, at least in the UK.

In terms of prompting strategies, previous experiments with ChatGPT prompts on this type of task have shown that they are relatively robust in the sense that ChatGPT works well on instructions intended for humans with only minor modifications and all attempts to improve or streamline the prompts so far have failed (Thelwall, 2025). Of course, this does not mean that better prompting strategies do not exist and future insights into how LLMs translate texts into score predictions might generate new prompting ideas.

## *Comparison with prior work*

The finding of a statistically significant (the 95% confidence interval excludes 0) positive article-level correlation between ChatGPT scores and departmental mean REF scores contrasts with the negative correlation previously found (Thelwall & Yaghi, 2024). From Figure 1, the reason for the negative correlation is that two of the departments chosen for the previous small-scale assessment (Leicester and Warwick) were anomalies against a general positive pattern. The results also contrast with a non-significant correlation found for ChatGPT 4o from a previous study of unpublished submissions to a medical journal (Saad et al., 2024), but this may have been due to a small sample size and the use of a single ChatGPT estimate per article.

## *Journal-based anomalies*

The results point to journal-based anomalies, so comparing extreme journals may reveal possible causes. Contrasting abstracts from NEJM and The Lancet with eLife, the main difference is that the former two journals are fact-based and therefore dry. The NEJM has structured abstracts, with the final section, Conclusions, used to summarise the results in non-statistical language without attempting to generalise or explicitly discuss their importance. For example, "Treatment with rosuvastatin at a dose of 10 mg per day resulted in a significantly lower risk of cardiovascular events than placebo in an intermediate-risk, ethnically diverse population without cardiovascular disease." In contrast, the Lancet's structured abstracts end with an Interpretation section that has the same role but seems to have more scope for tentative generalisation or wider context setting (e.g., "Functional status 90 days after intracerebral haemorrhage did not differ significantly between patients who received tranexamic acid and those who received placebo, despite a reduction in early deaths and serious adverse events. Larger randomised trials are needed to confirm or refute a clinically significant treatment effect.": Sprigg et al., 2018).  In contrast, eLife sometimes includes explicit claims for novelty and significance in its abstracts as well as general speculation about the importance of the results (e.g., "This raises the general concept that proteins involved in cytoskeletal functions and appearing organism-specific, may have highly divergent and cryptic orthologs in other species": Dean et al., 2019). Given this difference it seems possible that ChatGPT is influenced by the explicit strength of claims in abstracts, and REF reviewers may be influenced by points made in the main part of the paper, even if the difference is less important in other UoAs. Medical articles seem to be short (3000-4000 words) and should be understandable to medical practitioners, so REF experts might be more willing to read the full text, and authors might be more willing to omit relevant context from abstracts. Four examples in the Appendix illustrate this, with assertions that seem likely to influence ChatGPT in bold.

An alternative possibility is that ChatGPT does not consider the importance of human health when estimating medical research quality and so does not adequately assess the impact of the type of research typically found in The Lancet and NEJM. This could be a side-effect of medical articles not ever needing to spell out the importance of improving human health, so the significance of research is implicit rather than explicit. For example, the conclusion, "Triple antiplatelet therapy should not be used in routine clinical practice" (Bath, 2018) will reduce adverse side effects and, therefore, potentially save lives, although the abstract does not state this explicitly nor give an estimate of the number of lives that would be saved. A related issue is that originality for high quality medical research is arguably less important than rigour. For example, the same, or similar, intervention might be trialled in different populations (e.g., in different countries or patient groups) to determine its efficacy and cost-effectiveness. It seems unlikely that a medical expert would expect originality in research design given that there are hierarchies of evidence for health research (Evans, 2003) and a broad consensus over the best designs for important studies for human health, for example, randomised controlled trials, or systematic reviews of evidence. Similarly, an expert would be unlikely to criticise a study of a widely used treatment for a lack of originality if there was a reason why the study was needed. As suggested above, repeating a trial with a different population may lack originality but such studies are important because of the many small factors that can influence a well-designed study, and they can enable more precise effect size estimates when combined in meta-analyses (Schmid et al., 2020). As suggested above, it is also possible that REF panel experts attach more importance, and therefore weight, to the overall quality and potential impact of a research article, than the specific claims made in the abstract.

### Low scoring articles

The Word Association Thematic Analysis results for RQ4 suggested that ChatGPT favours theoretical studies and tends to give lower scores to studies directly informing human health decisions. To investigate this further, the lowest scoring individual articles were also examined to identify patterns. These tended to be patient studies with negative results. For example, the article with the lowest overall ChatGPT score concluded, "The MTD of sorafenib when used with 30 Gy in 10 fractions was not established due to sorafenib-related systemic toxicity. Severe radiotherapy-related toxicities were also observed. These events suggest this concurrent combination does not warrant further study" (Murray et al., 2017). This not only did not report positive results but also acknowledged unwanted side effects. Other examples of negative results from low scoring articles include, "Orally delivered PXD showed no evidence of clinical activity, when combined with weekly AUC2-carboplatin in PROC" (Fotopoulou et al., 2014) and "LY2495655 did not improve overall survival: the hazard ratio was 1.70 (90% confidence interval, 1.1–2.7) for 300 mg vs. placebo and 1.3 (0.82–2.1) for 100 mg vs. placebo (recommended doses)" (Golan et al., 2018).

Whilst it seems reasonable for ChatGPT to give lower scores to articles with negative findings since they are unlikely to have a "transformative effect" they may be important research contributions since other clinicians will not need to repeat the same study. For the same reason, studies with serious side effects may be particularly important to report. In addition, negative results about existing treatments can also be transformative if it leads to their discontinuation. It is not clear whether REF assessors took these wider issues into account, but it seems possible that negative clinical results are more likely to be published than negative theoretical results, and this would give an overall negative bias towards clinical studies in ChatGPT. Overall, this suggests that different (human expert) standards might apply to medical research involving patients or other participants, but that ChatGPT has not recognised this.

## Conclusion

The results show for the first time that ChatGPT score estimates correlate positively with an article quality proxy (departmental mean REF scores) for clinical medicine journal articles. This is true for ChatGPT 3.5 turbo, 4o-mini, and 4o, all with the same system prompts and also for ChatGPT 4o-mini with draconian prompts. These multiple tests with positive results therefore robustly override the previous small-scale test of the same field that had negative results (Thelwall & Yaghi, 2024). The practical value of this is that ChatGPT score estimates could be used to support human judgements in contexts that citation data currently is, including for clinical medicine articles that are too new to have sufficiently mature citation data (e.g., published in the previous two years). For the best estimates, at least five iterations of ChatGPT should be used for each article and a transformation should be applied to compensate for ChatGPT's positive bias.

As for citation-based indicators, ChatGPT score estimates should not replace human judgments for important decisions but instead support them (e.g., acting as a tie-breaker) or be used in contexts where it is impractical to commission expert evaluations of the articles due to a lack of available expertise or when there are too many articles to be evaluated (e.g., national or departmental evaluations). For example, the citation data for the US Science & Engineering Indicators report 2024 stopped in 2022, but if ChatGPT scores had been used instead, then 2023 might have been included for more recent evidence, and the results might be more accurate as a quality indicator. Similarly, the UK's Life Sciences Competitiveness Indicators 2024 (Gov.uk, 2024) could have used ChatGPT quality scores as a more accurate indicator of life science (including medicine) research quality. When ChatGPT scores are used, the same caution should be employed as for the citation-based indicators that they replace (Hicks et al., 2015).

The main caveat to the above conclusion is that ChatGPT seems to systematically undervalue research in some prestigious medical journals, whether because of the journal's style or by overlooking the contribution of medical studies about human health. This should be considered when using ChatGPT estimates, such as by estimating the quality of such articles in other ways or by not using ChatGPT-based indicators for them. Of course the citation data currently used in many tasks also has biases.

Finally, the results, combined with those of a previous study (Thelwall & Yaghi, 2024), give evidence that ChatGPT quality scores correlate positively with REF quality scores in all 34 UoAs, making it a universal quality indicator, albeit with its strength varying between fields.

# Acknowledgement

This study is funded by the Economic and Social Research Council.

## Appendix: Examples of ChatGPT scores for three medical journals

A NEJM article that received a low score from ChatGPT was, "Effects of serelaxin in patients with acute heart failure", with an abstract that concluded, "In this trial involving patients who were hospitalized for acute heart failure, an infusion of serelaxin did not result in a lower incidence of death from cardiovascular causes at 180 days or worsening heart failure at 5 days than placebo." One of the ChatGPT reports giving this article a score of 2* concluded, "The article provides a well-executed trial, adding to the corpus of literature on acute heart failure treatment modalities. However, it fails to present novel insights that would elevate its status to an internationally excellent or world-leading level. The findings reaffirm existing knowledge rather than challenge or expand it significantly. Suggestions for future research could include exploration of different dosages, combination treatments with serelaxin, or investigations into specific patient subgroups that may respond differently to treatment. Additional focus on long-term cardiovascular outcomes or quality of life measurements post-treatment could further enhance the study's impact and significance within the context of heart failure management."

An article in The Lancet with a low score from ChatGPT was, "Antiplatelet therapy with aspirin, clopidogrel, and dipyridamole versus clopidogrel alone or aspirin and dipyridamole in patients with acute cerebral ischaemia (TARDIS): a randomised, open-label, phase 3 superiority trial" which concluded, "Interpretation: Among patients with recent cerebral ischaemia, intensive antiplatelet therapy did not reduce the incidence and severity of recurrent stroke or TIA, but did significantly increase the risk of major bleeding. Triple antiplatelet therapy should not be used in routine clinical practice." A ChatGPT report recommending a 3* score stated, "In conclusion, the article presents compelling evidence against the routine use of triple antiplatelet therapy in patients after acute cerebral ischaemia, and while its originality, significance, and rigour are recognized as excellent, it stops short of achieving the highest standards of excellence. For improvement, the authors could further explore long-term outcomes and potential variations in patient subgroups, which could enrich understanding and yield further insights into patient management strategies."

An article from eLife with a high ChatGPT score (4*) was, "Curvature-induced expulsion of actomyosin bundles during cytokinetic ring contraction", with an abstract that concluded, "**Strikingly**, mechanical compression of actomyosin rings results in expulsion of bundles predominantly at regions of high curvature. **Our work unprecedentedly reveals** that the increased curvature of the ring itself promotes its disassembly. It is likely that such a curvature-induced mechanism may operate in disassembly of other contractile networks." ChatGPT's summary of one of its 4* evaluations was, "Overall, this article stands out as a world-leading contribution to the field, exemplifying high standards of originality, significance, and academic rigour. Its findings call for further exploration into the applications of PVCs [Photorhabdus Virulence Cassettes] not only in pest control but potentially in therapeutic areas involving human health and disease pathology as well. Future research could build on this work to investigate potential implications and applications in greater depth, reinforcing **the article's foundational contributions**." For context, "strikingly" has not appeared in the Lancet since 2013.

Another ChatGPT 4* article from eLife was, "A conserved major facilitator superfamily member orchestrates a subset of O-glycosylation to aid macrophage tissue invasion" which claimed in its abstract that, "We characterize **for the first time** the T and Tn glycoform O-glycoproteome of the Drosophila melanogaster embryo, and determine that Minerva increases the presence of T-antigen on proteins in pathways previously linked to cancer, most strongly on the sulfhydryl oxidase Qsox1 which we show is required for macrophage tissue entry. Minerva's vertebrate ortholog, MFSD1, rescues the minerva mutant's migration and T-antigen glycosylation defects. We **thus identify a key conserved regulator** that orchestrates O-glycosylation on a protein subset to activate a **program governing migration steps important for both development and cancer metastasis**."